\documentclass[12pt,preprint]{aastex}

\slugcomment{Submitted to The Astrophysical Journal}

\shorttitle{Magnetically Arrested Disks}
\shortauthors{I. V. Igumenshchev}

\begin{document}

\title{MAGNETICALLY ARRESTED DISKS AND\\ 
ORIGIN OF POYNTING JETS: NUMERICAL STUDY}

\author{Igor V. Igumenshchev}
\affil{Laboratory for Laser Energetics, University of Rochester\\
250 East River Road, Rochester, NY 14623}
\email{iigu@lle.rochester.edu}

\begin{abstract} 

The dynamics and structure of accretion disks, which accumulate the
vertical magnetic field in the centers, are investigated 
using two- and three-dimensional MHD simulations.
The central field can be built up to the equipartition level and disrupts
a nearly axisymmetric outer accretion disk inside a magnetospheric radius, 
forming a magnetically arrested disk (MAD).
In the MAD, the mass accretes in a form of irregular dense spiral streams
and the vertical field, split into separate bundles, 
penetrates through the disk plane in low-density magnetic islands.
The accreting mass, when spiraling inward, drags the field and
twists it around the axis of rotation,
resulting in collimated Poynting jets in the polar directions.
These jets are powered by the accretion flow with the efficiency 
up to $\sim 1.5\%$ (in units $\dot{M}_{in}c^2$).
The spiral flow pattern in the MAD 
is dominated by modes with low azimuthal wavenumbers $m\sim 1-5$ and
can be a source of quasi-periodic oscillations in the outgoing radiation.
The formation of MAD and Poynting jets can naturally explain the
observed changes of spectral states in Galactic black hole binaries.
Our study is focused on black hole accretion flows;
however, the results
can also be applicable to accretion disks around nonrelativistic
objects, such as young stellar objects and stars in binary systems.

\end{abstract}

\keywords{accretion, accretion disks --- black hole physics --- galaxies: 
jets --- gamma rays: bursts --- instabilities --- ISM: 
jets and outflows --- magnetic fields --- MHD --- turbulence}

\section{Introduction}

Accretion disks can carry small- and large-scale magnetic fields.
The small-scale field ($\ell\la R$, where $\ell$ is the field 
scale length and $R$ measures the radial distance from the disk center)
can be locally generated by the MHD dynamo
(Brandenburg et al. 1995; Stone et al. 1996) supported by
the turbulence, which results from the 
magneto-rotational instability (MRI, Balbus \& Hawley 1991).
This field can provide the outward transport of angular momentum
in the bulk of the disk with the help of local Maxwell stresses 
(Shakura \& Sunyaev 1973; Hawley, Gammie, \& Balbus 1996).
The large-scale field ($\ell > R$) is unlikely produced in 
accretion disks (however, see Tout \& Pringle 1996),
and can either be captured from the environment and 
dragged inward by an accretion flow (Bisnovatyi-Kogan \& Ruzmaikin 
1974, 1976), or inherited from the past evolution (see \S 4).
The large-scale field can remove the angular momentum from accretion disks
by global Maxwell stresses through the magnetized disk corona (K\"onigl 1989).

A large-scale bipolar field, unlike a small-scale field, 
can not dissipate locally due to the magnetic diffusivity
and can not be absorbed by the central black hole.
In the case of inefficient outward diffusion of the bipolar field
through the disk
(see Narayan, Igumenshchev, \& Abramowicz 2003; Spruit \& Uzdensky 2005; 
Bisnovatyi-Kogan \& Lovelace 2007; and
for other possibility, see van Ballegooijen 1989;
Lubow, Papaloizou, \& Pringle 1994; 
Lovelace, Romanova, \& Newman 1994; Heyvaerts, Priest, \& Bardou 1996;
Agapitou \& Papaloizou 1996; Livio, Ogilvie, \& Pringle 1999),
this field is accumulated in the innermost region of an accretion disk
and forms a ``magnetically arrested disk,'' or MAD 
(Narayan et al. 2003).
The MAD consists of two parts: the outer, almost axisymmetric, 
Keplerian accretion disk and
the inner magnetically dominated region, in which the accumulated
vertical field disrupts the outer disk at the magnetospheric
radius $R_{\rm m}\sim 8\pi GM\rho/B^2$, where $M$ is the central mass,
$\rho$ is the accretion mass density, and $B$ is the magnetic induction.

It is believed that the large-scale bipolar field in accretion disks
is responsible for the formation of jets
observed in a large variety of astrophysical objects
(e.g., Livio, Pringle, \& King 2003).
The magnetically driven jets can be of two types,
basically depending on a mass load by the disk matter
(e.g., Lovelace, Gandhi, \& Romanova 2005):
Poynting jets and hydromagnetic jets, which
have, respectively, small and large mass loads.
The hydromagnetic jets can be formed by two mechanisms:
the magneto-centrifugal mechanism 
(Blandford \& Payne 1982; K\"onigl \& Pudritz 2000) 
and the toroidal-field pressure generated by the disk rotation
(Lynden-Bell 2003; Kato, Mineshige, \& Shibata 2004).
The magneto-centrifugal mechanism
produces relatively wide outflows and, to be consistent with observations
of the collimated jets, requires an additional focusing mechanism. 
Jets driven by the toroidal-field pressure can have
a high degree of collimation, but these jets are known to be kink 
unstable (Eichler 1993; Appl 1996; Spruit, Foglizzo, \& Stehle 1997).
Poynting jets  are naturally self-collimated and
expected to be marginally kink stable
(Li 2000; Tomimatsu, Matsuoka, \& Takahashi 2001).
These jets can originate in the innermost region of accretion disks
and powered either by the disks themselves (Lovelace, Wang, \& Sulkanen 1987;
Lovelace et al. 2002)
or by rotating black holes (Blandford \& Znajek 1977; Punsly 2001;
also see, 
Takahashi et al. 1990; Komissarov 2005; Hawley \& Krolik 2006; McKinney 2006).

Our study is based on two- and three-dimensional
(2D and 3D, respectively) MHD simulations
and has two main goals.
First, we investigate the dynamics and structure of MAD's.
We show that MAD's are formed in the accretion flows, which carry inward
large-scale poloidal magnetic fields. 
Inside the magnetospheric radius $R_{\rm m}$, 
the matter accretes as discrete streams and blobs, fighting its way
through the strong vertical magnetic field fragmented in separate bundles.
Because of rotation, the streams take spiral shapes.
Second, we demonstrate a link between the existence of MAD's and production
of powerful Poynting jets.
These jets should always be generated in MAD's
because of the interaction of the spiraling accretion flow
with the vertical magnetic bundles, which, as the result, are twisted around
the axis of rotation.

This paper extends the work of Igumenshchev, Narayan, \& Abramowicz (2003)
by studying in more detail the radiatively inefficient accretion disks 
with poloidal magnetic fields.
We employ a new version of our 3D MHD code, which can be utilized in
multi-processor simulations.
The paper is organized as follows:
We describe the solved equations, the numerical method used, and 
initial and boundary conditions in \S 2.
We present our numerical results in \S 3, and
discuss and summarize them in \S 4.

\section{Numerical method}

We simulate nonradiative accretion flows around 
a Schwarzschild black hole of mass $M$ using the following equations
of ideal MHD:
\begin{equation}
{d\rho\over dt} + \rho{\bf\nabla\cdot v} = 0,
\end{equation}
\begin{equation}
\rho{d{\bf v}\over dt} = -{\bf\nabla} P - \rho{\bf\nabla}\Phi +
{1\over 4\pi}({\bf\nabla}\times{\bf B})\times {\bf B},
\end{equation}
\begin{equation}
{\partial\over\partial t}\left(\rho{v^2\over 2}+\rho\epsilon+
{B^2\over 8\pi}+\Phi\right)=-\nabla\cdot{\bf q},
\end{equation}
\begin{equation}
{\partial{\bf B}\over \partial t} = {\bf\nabla}\times({\bf v}\times{\bf B}),
\end{equation}
where 
${\bf v}$ is the velocity, $P$ is the
gas pressure, $\Phi$ is the gravitational potential, 
$\epsilon$ is the specific internal energy of gas, 
and ${\bf q}$ is the total energy flux per unit square 
(see, e.g., Landau \& Lifshitz 1987).
We adopt the ideal gas equation of state,
\begin{equation}
P=(\gamma-1)\rho\epsilon,
\end{equation}
with an adiabatic index $\gamma=5/3$.
We neglect self-gravity of the gas 
and employ a pseudo-Newtonian approximation (Paczy\'nski \&
Wiita 1980) for the black hole potential
\begin{equation}
\Phi=-{GM\over R-R_g},
\end{equation}
where $R_g=2GM/c^2$ is the gravitational radius of the black hole.
No explicit resistivity and viscosity are applied in equations (2)-(4).
However, because of the use of the total energy equation (3), 
the energy released
due to the numerical resistivity and viscosity is consistently accounted
as heat in our simulations. 

The MHD equations (1)-(4) are solved employing the time-explicit
Eulerian finite-difference method, which is an extension
to MHD of the hydrodynamic piecewise-parabolic method by Colella \&
Woodward (1984). 
We solve the induction equation (4) using the constrained transport
(Evans \& Hawley 1988; Gardiner \& Stone 2005), which preserves 
the $\nabla\cdot{\bf B}=0$ condition.
In our method, we employ the approximate MHD Riemann solver by Li (2005).
Test simulations have shown that this solver is robust and provides
a good material interface tracking.

We use the spherical coordinates $(R,\theta,\phi)$. 
Our 3D numerical grid has $182\times 84\times 240$
zones in the radial, polar, and azimuthal directions, respectively. 
The radial zones are spaced logarithmically from $R_{in}=2 R_{g}$
to $R_{out}=220 R_{g}$.
Both hemispheres are considered, in which polar cones with the opening
angle $\pi/8$ are excluded. Therefore, the polar domain extends
from $\theta=\pi/16$ to $15\pi/16$.
The grid resolution in the polar direction is gradually changed from
a fine resolution around the
equatorial plane to a coarse resolution near the poles 
(with the maximum-to-minimum grid-size ratio $\approx 3$).
The azimuthal zones are uniform and cover the full $2\pi$ range
in $\phi$.
The absorption condition for the mass and 
the condition of the zero-transverse magnetic field
are applied in the inner and outer radial boundaries,
providing that the mass and field can freely leave the computational
domain through these boundaries.
In the boundaries around the excluded polar regions, 
we apply the reflection boundary conditions,
which means that no streamlines and magnetic lines can go through
these boundaries.

At the beginning of our simulations, 
the computational domain is filled with a very low-density,
nonmagnetized material.
The simulations are started in 2D, assuming the axial symmetry,
with an injection of mass in a slender torus, which is
located in the equatorial plane at $R_{inj}=210\, R_{g}$;
i.e., close enough to $R_{out}$.
This mass has the Keplerian angular momentum and specific internal energy
$\epsilon_{inj}=0.045\,GM/R_{inj}$.
After an initial period of simulation 
without magnetic fields, the mass forms a steady
thick torus, which has the inner edge at $R\approx 150\,R_{g}$ and
which outer half is truncated at $R_{out}$.
The torus contains a constant amount of mass and is in a dynamic equilibrium:
all the injected mass flows outward through
$R_{\rm out}$ after a circulation inside the torus.
No accretion flow at this point is formed.
This hydrodynamic, steady, thick torus is used as an initial configuration
in our MHD simulations.

The MHD simulations are started at $t=0$ from the steady, thick torus
by initiating the injection of a poloidal magnetic field 
into the injection slender torus at $R_{inj}$.
The numerical procedure for the field injection is
similar to that described by Igumenshchev et al. (2003) 
except for one modification:
now the strength of the injected field 
can be limited by setting the minimum $\beta_{\rm inj}$, 
which is the ratio of the gas pressure to 
the magnetic pressure at $R_{\rm inj}$.
This modification allows us to better control the rate of field injection.
The entire volume of the thick torus is filled by the field during 
about one orbital period, $t_{\rm orb}$, estimated at $R_{\rm inj}$. 
Since this moment, $t\simeq t_{orb}$, the formation of accretion flow begins
as a result of redistribution of the angular momentum in the torus due to
Maxwell stresses.
In the following discussion, we will use the time normalized by
the time-scale $t_{orb}$, i.e. $t\rightarrow t/t_{\rm orb}$.

\section{Results}

We present the results of combined axisymmetric 2D and 
non-axisymmetric 3D simulations. The initial evolution
in our models has been simulated in 2D.
This allows us to consider longer evolution times
in comparison to those that can be obtained in 3D simulations, 
because of the larger requirements for computational resources 
in the latter case.
We initiate 3D simulations starting from developed
axisymmetric models. The results 
have shown that non-axisymmetric motions are not very important
in the outer parts of the constructed accretion flows and, therefore,
the use of 2D simulations on the initial evolution stages is
the reasonable simplification.

We consider three models, which differ by
the rates of field injection determined by
$\beta_{\rm inj}=10$, 100, and 1000, and we will refer to these models
as Model~A, B, and C, respectively. All other properties of the models,
including the injection radius $R_{inj}$, internal energy $\epsilon_{inj}$,
and numerical resolution, are the same (see \S 2).

\subsection{Accretion flows}

The initial axisymmetric development of the models is
qualitatively similar: the inner edge of the thick torus
is extended toward the black hole, forming relatively
thin, almost Keplerian accretion disks. 
The time of the disk development
is varied, depending on the strength of the injected field.
The accretion of the mass into the black hole begins at
$t\approx 0.7$ in Model~A,
$\approx 1.3$ in Model~B, and $\approx 4.2$ in Model~C
(time is given in units of the orbital period at $R_{inj}$, see \S 2).
At this stage, the evolution of the
disks is governed mainly by global Maxwell stresses
produced by the poloidal field component.
This component is advected inward with the accretion flow and,
because of the disks' Keplerian rotation,
generates relatively strong toroidal magnetic fields localized above and 
below the mid-plane.
These toroidal fields form a highly magnetized disk corona with a typical
$\beta\sim 0.01$. Model~B and, especially Model~C, demonstrate
the development of 2D MRI.
This development is similar to that
observed by Stone \& Pringle (2001); in particular, in their ``Run C."
We have found the origin of the
channel solution (see Hawley \& Balbus 1992)
in the central regions of Models~B and C.
This solution consists of oppositely directed radial streams
and is the characteristic feature of the axisymmetric
non-linear MRI (Stone \& Norman 1994).
Analysis of the models shows that the channel solution is developed when the 
wavelength $\lambda=2\pi V_A/\sqrt{3}\Omega$ 
of the fastest growing mode of the MRI is
well resolved on the numerical grid, 
i.e. $\lambda\ga 5\Delta x$, where $\Delta x$
is the grid size, $V_A$ is the Alfv\'en velocity, and $\Omega$ is
the angular velocity.
Model~A shows no indication of the MRI, which can be attributed 
to the strong magnetic fields, which suppress the instability.
In this model, the estimate of $\lambda$ typically exceeds 
the disk thickness.
Model~A has some resemblance to nonturbulent
``Run F'' of Stone \& Pringle (2001).
In spite of the mentioned similarities with the results of
Stone \& Pringle (2001),
our models show different behavior on the long evolution times.
Our simulation design with
the permanent injection of mass and magnetic field results in
accretion disks, which accumulate the poloidal field
in the center and form MAD's. The models of Stone \& Pringle (2001; also
De Villiers, Hawley, \& Krolik 2003; Hirose et al. 2004;
McKinney \& Gammie 2004; Hawley \& Krolik 2006) 
did not form MAD's and did not show a long-time
accretion history, probably because of the initiation of simulations from
static magnetized tori, which contain
a limited amount of mass and magnetic flux of one sign. 
We will concentrate on the results describing the formation, evolution, 
and structure of MAD's in the following text.
Other aspects of our results will be reported elsewhere.

Figure~1 shows example snapshots of the axisymmetric density distribution in
Model~B from 2D simulations at two successive moments, 
$t=5.1153$ and $5.1458$.
The accretion flow is nonuniform because of the development of turbulence. 
The turbulence results from the combined effect
of the MRI and current sheet instability. The latter instability locally
releases heat due to reconnections of the oppositely directed toroidal
magnetic fields.
The reconnection heat makes a significant contribution to the local energy
balance in the central regions of the flow,
because of the relatively high energy density of the field, which
is comparable to the gravitational energy density of the accretion mass.
The thick disk structure observed in Fig.~1 is explained by
convection motions supported by the reconnection heat.
Note that the case, in which the turbulence is
supported by only convection motions
from the reconnections,
without the effects of rotation and MRI, 
had been demonstrated in simulations of spherical magnetized
accretion flows (Igumenshchev 2006).
In the case of disk accretion,
almost axisymmetric convection motions, similar to those found here,
had been observed in 3D models
with toroidal magnetic fields (Igumenshchev et al. 2003).
The convection motions in Models~B and C make 
these models relevant to
convection-dominated accretion flows (Narayan, Igumenshchev, \& 
Abramowicz 2000; Quataert \& Gruzinov 2000).
Our simulations show that the poloidal 
field is transported inward in axisymmetric turbulent flows 
and accumulated in the vicinity of the black holes. When the central poloidal
field reaches some certain strength (about equipartition with
the gravitational energy of the accreting mass), the accretion flow 
becomes unstable (Narayan et al. 2003).
In axisymmetric simulations, the instability takes the form of cycle
accretion, in which the more-extended periods of halted accretion 
(see Fig.~1a) are followed by
the relatively short periods of accretion (see Fig.~1b).
In the case of the halted accretion period, the inner accretion disk is
truncated at the magnetospheric radius $R_{m}$, which is
$\approx 15\,R_g$ in Fig.~1a.
The pressure of the strong central vertical field (see Fig.~2a) prevents 
the mass accumulated at $R_{m}$ from falling into the black hole.
The accretion begins as soon as the gravity of 
the accumulated mass overcomes the magnetic pressure.
During the accretion period, the whole magnetic flux,
which is localized inside $R_{\rm m}$ in the
period of halted accretion, is moved on the black hole horizon
(see Fig.~2b). Note that similar structural features of the inner MHD
flows in accretion disks related to the model of gamma-ray bursts
were discussed by Proga \& Zhang (2006).

Figure~3 illustrates the time dependence of the accretion flow in
Model~B, showing the evolution of the mass accretion rate $\dot{M}_{in}$
and magnetic fluxes
in the midplane inside the five specific radii: $210\,R_g$ 
($=R_{inj}$), $100\, R_g$, $50\, R_g$, $25\, R_g$, 
and $2\, R_g$ ($=R_{inj}$).
This figure shows the evolution, which has been simulated 
in 2D from $t=0$ to $2.14$ 
and in 3D after $t=2.14$.
The vertical dashed line in Fig.~3 indicates the moment of initiation
of the 3D simulations.
The cycle accretion in the 2D simulations begins at $t\approx 1.4$
and is clearly seen as a sequence of spikes 
in the time-dependence of $\dot{M}_{in}$ (see Fig.~3a).
Spikes, which are related to the same cycle
accretion, are also observed in the variation of magnetic flux inside
$R=2\, R_g$ (see Fig.~3b). 
The magnetic fluxes inside the other selected radii are gradually
increased with time because of the inward advection of the vertical field. 
The time dependence of these fluxes is not significantly influenced 
by the cycle accretion.

The structure and dynamics of the inner region in Model~B are drastically
changed in the 3D simulations.
Shortly after the initiation of the 3D simulations at $t=2.14$,
the axisymmetric distribution of mass near $R_{\rm m}$
undergoes the Rayleigh-Taylor and, possibly,
Kelvin-Helmholtz instability (see Kaisig, Tajima, \& Lovelace 1992;
Spruit, Stehle, \& Papaloizou 1995;
Chandran 2001; Li \& Narayan 2004)
with the fastest growing azimuthal mode number $m\simeq 50$
(the latter is probably determined by our grid resolution).
As a result,
the empty region inside $R_{\rm m}$ is quickly filled,
on the free-fall time scale estimated at $R_{\rm m}$, with the large
number of density spikes moved almost radially toward the center.
These spikes quickly disappear in the black hole and, at a later time,
the inner disk structure is modified toward establishing a 
dynamic quasi-steady state. This state is characterized by
a low $m$-mode ($m\simeq 1$-5) spiral-flow structure,
which results from the interaction of the accreting mass with
the strong vertical magnetic field.
Note that the similar low $m$-mode flow structure
was found in the simulations
of accretion flows onto a magnetic dipole (Romanova \& Lovelace 2006).

The non-axisymmetric inner flow is highly time-variable and experiences a 
quasi-periodic behavior.
Figure~4 shows an example of the flow structure inside 50 $R_g$
in Model~B, at two successive moments: $t=2.2767$ and 2.2867.
The flow is essentially 3D inside the magnetically
dominated region limited by the radius $\simeq 35\,R_g$
and remains almost axisymmetric on the outside of this radius.
In the magnetically dominated region, 
the flow forms moderately tightened spirals of dense
matter, which are clearly seen in Figs~4a and 4b. 
This matter is quickly
accreted into the black hole with the radial velocity, which is
$\sim 0.5$ a fraction of the free-fall velocity.
Such a relatively fast infall is explained by the efficient loss
of the angular momentum by the mass during its interaction with
the vertical field. 
The field is distributed nonuniformly in the disk plane, 
concentrating in bundles that penetrate through
the plane in very low density,
magnetically dominated (with $\beta\sim 0.01$) regions, 
or magnetic ``islands.''
The rotating mass interacts with magnetic bundles
and forces them to twist around the disk's rotational axis.
In the simulations, this twist is observed as the rotation of
magnetic islands around the center in the disk plane.
The rotational velocity of the islands typically has the
reduced rotational velocity by the factor of $\sim0.5-1$
in comparison with the velocity of the surrounding accretion matter.
This can be explained by the resistance
of the large-scale vertical field to such a twist.
The faster rotation of accretion matter and
slower rotation of magnetic islands produces a shear flow.
The shear flow plays two roles in our simulations.
First, it provides the exchange of momentum and energy between
the accreting mass and vertical field.
Second, the shear flow
results in an ablation of the islands caused by magnetic
diffusivity, making each island a temporal structure.
An example evolution of magnetic islands can be seen in Fig.~4:
the magnetic islands observed as low-density spiral arms 
above the center in Fig.~4a are observed below the center in Fig.~4b,
after about half a revolution in the clockwise direction.
In the latter figure,
the islands are apparently reduced in size due to the ablation.

The vertical field ablated from magnetic islands
is carried inward by the accretion flow and accumulates on the 
black hole horizon. This accumulation results in
quasi-periodic eruptions of the field outward from the horizon 
as soon as the field pressure overcomes
the dynamic pressure of the accreting mass.
The eruptions typically take the form of high-velocity narrow streams
(in the equatorial cross-section) of a low-density, 
magnetically dominated medium fountained outward from the black hole. 
In Fig.~4b, four magnetic islands observed as low-density regions 
inside $R\approx 15\, R_g$
result from such eruptions and the eruption of one of these islands
(to the right from the center; see also the steam that produced it) 
still continues at the moment shown.
In the consequent evolution, these islands are pushed outward 
and stretched in the azimuthal direction by the accretion flow, and
take the spiral shape similar to that shown in Fig.~4a.

Model~A evolves faster and accumulates a larger magnetic flux at the
center in comparison with Model~B.
Figure~5 shows the evolution of the accretion rate $\dot{M}_{\rm in}$
and magnetic fluxes in Model~A.
Qualitatively, the evolution of these quantities is similar to 
the evolution of those in Model~B (see Fig.~3). 
Quantitatively, however, Model~A demonstrates
significantly larger 
time-averaged accretion rates (by about two orders of magnitude) and 
longer quasi-periods of the cycle accretion
(represented by the intervals between spikes
in the time-dependence of $\dot{M}_{\rm in}$ in Fig.~5a)
in the 2D simulations. 
By the end of the 2D simulations at $t\approx 1.7$, this model has 
the maximum $R_m\simeq 30-40\,R_g$.

In the 3D simulations,
Model~A experiences the initial transient period, 
similar to the period of the development of
the Rayleigh-Taylor instability in Model~B (see above), in which
the non-axisymmetric, small-scale structures quickly appear
and disappear.
Figure~6 shows an example of the developed low $m$-mode spiral structure 
in Model~A obtained after the transient period.
This structure is clearly dominated by the $m=1$ mode.
The magnetically dominated region is extended up to
$R\simeq 70\, R_g$.
Note that the spiral-density arms seen in Fig.~6 
are more open than the arms in Model~B
(see Fig.~4a). This could be due to the stronger central field
in Model~A.

Model~C is our slowest evolving model and,
accordingly, shows the slowest rate of accumulation 
of the central vertical field.
This model has been calculated only in 2D and
demonstrated the qualitative similarity to the axisymmetric 
evolution of Models~A and B.
The cycle accretion, which is caused by the accumulated field,
begins at $t\approx 4.8$ in Model~C.
The model demonstrates more-efficient turbulent motions in the
accretion flow.
This can be attributed to weaker magnetic fields, which
suppress less the MRI
and convection motions. At the end of simulation
at $t\approx 6$, the model has the maximum $R_m\simeq 6\, R_g$.

\subsection{Poynting Jets}

The 3D simulations of Models~A and B
show that the vertical field penetrated the central
magnetically dominated regions in MAD's is twisted around
the axis of rotation by the rotating accretion flows.
The field twist generates electromagnetic perturbations, which
propagate outward and transport the released energy
in the form of a Poynting flux (e.g., Landau \& Lifshitz 1987)
\begin{equation}
{\bf S}={1\over 8\pi}(({\bf v}\times{\bf B})\times{\bf B}).
\end{equation}
The Poynting flux is distributed nonuniformly in the polar angles,
basically showing two components:
a jet-like concentration of the flux near the poles and
a wide-spread distribution of the flux in the equatorial and
mid-polar-angle directions (from $\theta\sim\pi/4$ to $\sim 3\pi/4$).

Figures~7 and 8 show example $\theta$-distributions of the 
radial Poynting flux (solid lines),
\begin{equation}
S_R=v_R{B^2\over 4\pi}-{B_R\over 4\pi}({\bf v}\cdot {\bf B}),
\end{equation}
at six different radii, 5 $R_g$, 10 $R_g$, 25 $R_g$, 50 $R_g$,
100 $R_g$, and 220 $R_g$,
which cover most of the radial domain
in Models~B and A, respectively. The shown distributions are averaged in
the azimuthal direction and in time over the interval $\Delta t\simeq 0.05$, 
using a set of data files stored during the simulations.
Outside of the
inner magnetically dominated region (at $R\ga 35\,R_g$ in Model~B and 
$R\ga 70\,R_g$ in Model~A),
the outward (positive) Poynting flux in the equatorial and 
mid-polar-angle directions
is supported mostly by the rotation of the outer,
almost axisymmetric disks, in which the poloidal field component is frozen in.
Such a flux is present in both the axisymmetric 2D and 3D simulations.
The 3D simulations introduce new important features in
the Poynting flux distribution:
an increase of the flux at the equatorial and mid-polar-angle
directions inside the magnetically dominated region, and 
at the polar directions outside this region (see Figs~7 and 8).
The equatorial flux inside the magnetically dominated region
is generated due to the twist of the vertical field by the spiraling 
non-axisymmetric accretion flows. In Model~B, this flux,
represented by bumps in the $\theta$-distributions, gradually
deviates from the equatorial plane toward the poles as the radial
distance increases (see Figs~7a-7d). At $R\ga 100\,R_g$, the flux
is collimated into bi-polar Poynting jets
(see Figs~7e and 7f).
In the case of Model~A, the process of jet collimation
is less evident and somewhat different; but still, one can observe the
formation of narrow bi-polar Poynting jets starting from $R\simeq 10\,R_g$
and further development of these jets at large radii (see Figs~8b-8f).

Figures~7 and 8 show $\theta$-distributions of the kinetic
flux (dashed lines),
\begin{equation}
F_R=v_R\rho{v^2\over 2},
\end{equation}
for comparison with the Poynting flux. 
Typically, the kinetic flux is comparable, 
but does not exceed the Poynting flux in the polar jets 
(except in the outermost region in Model~A, see Fig.8f). 
Accordingly, the jet velocity is mostly sub-Alfv\'enic.
However, the value of the kinetic flux is relatively large and 
this is in some contradiction
with our expectations that MAD's can develop Poynting flux dominated jets.
The problem of the excessive kinetic flux in our simulations
can probably be explained
by the action of the numerical magnetic diffusivity (see \S 4), 
which results in an unphysically large mass load of the Poynting jets 
and the consequent excessive kinetic flux in them.

The Poynting jets are powered by 
the released binding energy of the accretion mass.
To estimate quantitatively the amount of energy going into the jets,
we calculate the Poynting jet ``luminosity" $\dot{E}_{jet}$
as a function of the radius R,
\begin{equation}
\dot{E}_{jet}(R)=\int R^2 S_R\,d\Omega,
\end{equation}
where 
the integration is taken over the solid angles $\Omega$
occupying the polar regions with $\theta < \pi/4$
and $\theta > 3\pi/4$ (excluding the boundary polar cones, see \S 2).
For comparison, we also calculate the Poynting total luminosity 
$\dot{E}_{tot}$, which is defined analogously to $\dot{E}_{jet}$,
but with the integration in eq.~(9) taken over the whole sphere.
Figures~9 and 10 show the radial profiles of the normalized 
$\dot{E}_{jet}$ (solid lines) and $\dot{E}_{tot}$ (dashed lines)
in Models~B and A, respectively.
The jet luminosity $\dot{E}_{jet}$ weakly depends on the radius
at $R\ga 50\,R_g$ in both models and equals to
$\approx 1.5\%$ in Model~B and $\approx 0.5\%$ in Model~A.
Here, we quantify the luminosity in the units of 
accretion power $\dot{M}_{in}c^2$.
The total luminosity $\dot{E}_{tot}$ includes the flux from
the bi-polar Poynting jets and wide equatorial Poynting outflow.
The latter component of $\dot{E}_{tot}$ exceeds $\dot{E}_{jet}$
by the factor of $\sim 3$ at large radii (see Figs~9 and 10).
This can be the consequence of the employed simulation design
(see \S 2), in which
the disk accretion at outer radii
is mostly provided by the global Maxwell stresses.

The smaller value of the final (at large radii)
relative $\dot{E}_{jet}$ in Model~A, 
in comparison with that in Model~B, 
can be explained by the different structure of the
inner magnetically dominated region in these models.
Model~A has the less tighten spiral density arms (see Fig.~6), 
in which the mass accretes with larger radial velocity and, therefore,
delivers less energy to the field.
Note, also, that the value of $\dot{E}_{jet}$ 
in Model~A takes the relatively large
finite value right at the inner boundary $R_{in}$ 
(see Fig.~10), whereas $\dot{E}_{jet}$
in Model~B begins from a small value at $R_{in}$ 
and gradually increases outward (see Fig.~9). 
This difference in the behavior of $\dot{E}_{jet}$
can be attributed to the discussed difference of the innermost structure
in the considered models.

\section{Discussion and Conclusions}

We have performed a numerical study of the formation and
evolution of quasi-stationary MAD, which is
characterized by a strong vertical magnetic field
accumulated at the disk center.
We employ the simulation design, in which the 
poloidal magnetic field of one sign is permanently injected into the
computational domain at the outer numerical boundary and
the unipolar vertical field is transported inward by the accretion flow.
The accumulated field has a significant impact on the inner flow structure and
dynamics in both 2D and 3D simulations.
In the axisymmetric 2D simulations, 
the field pressure can temporarily halt
the mass falling into the black hole, resulting in
the cycle accretion, in which the longer periods of accumulation
of the mass at the magnetospheric radius $R_m$ 
are followed by the short periods of accretion.
The 3D simulations have shown, however, that the axisymmetric cycle accretion
is not realized.
Instead, the accumulated field
causes the mass to accrete quasi-regularly in the form of non-axisymmetric
spiral streams and blobs.
We have demonstrated that 3D MAD's can be efficient sources of
collimated, bipolar Poynting jets, 
which originate in the vicinity of the central black hole. 
These jets develop due to and are powered by
the interaction of the spiral mass inflows
with the central field split into separate magnetic bundles.
The efficiency of conversion of the accretion energy $\dot{M}_{in} c^2$
into the Poynting jet energy is up to 1.5\% in our simulations. 
This estimate may not be accurate
(we believe, underestimated) because of the use
of the pseudo-Newtonian approximation (see \S 2).
The better estimate of the efficiency can be obtained using
general relativistic MHD simulations.

We have presented the simulation results from
three models of radiatively inefficient accretion disks, 
which differ by the strength of 
the injected field. In accordance with the previous studies
(e.g., Stone \& Pringle 2001), the structure of the outer disks
in these models is determined by the field strength.
In Model~A, which has the largest injected field, 
the MRI is suppressed and the accretion flow is driven by
global Maxwell stresses, which transport
the excessive angular momentum outward
from the disk through the disk corona. 
In Models~B and C, which have the smaller injected fields, 
the MRI and turbulent motions are developed.
The turbulence in these models is axisymmetric and 
partially supported by the efficient convection motions resulting from
dissipation of toroidal and small-scale, poloidal magnetic fields.
These motions cause the increase of the disk thicknesses
in comparison with non-turbulent Model~A.
The 3D simulations of Models~A and B have demonstrated that 
non-axisymmetric motions are not important in the outer parts of the disks, 
outside the inner magnetically dominated region (the disks remain
almost axisymmetric), 
but very important inside this region,
resulting in the development of the spiral accretion flows
and bi-polar Poynting jets.

Numerical magnetic dissipations and reconnections
result in a magnetic diffusivity, which
influences the structure and dynamics of our models.
The spatial scale, on which the diffusivity
occurs in our simulations, 
is determined by the gridsize, which greatly exceeds
the scales of various resistive mechanisms (e.g., Coulomb collisions,
dissipative plasma instabilities)
in the relevant astrophysical conditions.
Therefore, the magnetic diffusivity is significantly overestimated
in our models and results in an excessive
slippage of an accretion flow through magnetic field.
This slippage reduces the ability of the flow to drug
inward the vertical field, but, however, the numerical diffusivity
is not efficient enough
to totally prevent the field accumulation at the disk center.
The numerical diffusivity suppresses the MRI on the scales
of the gridsize, and, therefore, prevents
the development of turbulence in the outer regions of our models,
where the gridsize is increased.
Other effect of the numerical magnetic diffusivity is 
the enhancement of the ablation of
magnetic islands, which are found in the 3D simulations (see \S 3.1).
To test the sensitivity of our models to 
magnetic dissipations,
we have performed a 2D simulation of the model, which is
similar to Model~A, but has the double number of grid points in 
the $R$- and $\theta$-directions. 
The simulation has demonstrated the qualitative similarity of 
axisymmetric evolution 
of the high resolution model and Model~A:
both models show the formation of accretion disks,
accumulation of the vertical field in the disk centers, and
development of the cycle accretion. 
The high resolution model forms a thiner laminar accretion disk.
Unfortunately, a more detailed quantitative
comparison of these models meets some difficulties because of
the different properties of the mass
and field injection region (see \S 2),
which are changed with the change of the resolution.

The main results of our study, 
the formation of MAD's and Poynting jets,
have been obtained under the assumption of radiatively inefficient flows,
but, we believe that these results
can also be applied to the radiatively efficient, dense
accretion disks (e.g., Kato, Fukue, \& Mineshige 1998).
The formation of MAD's should not be affected by the 
radiative cooling as soon as the central field satisfies 
the equipartition condition,
\begin{equation}
{B^2\over 8\pi} \sim {GM\rho\over R_g},
\end{equation}
where $\rho$ is the mass density in the innermost region.
The radiative losses results in the higher $\rho$ and, therefore, 
the larger B is necessary to obtain radiative MAD's.
We expect that the qualitatively similar spiral
structure of the inner magnetically dominated region, 
to that found in our simulations, can be developed
in the case of the radiative disks.
Poynting jets should be a necessary attribute of the radiative MAD's as well.

The formation of MAD's in the radiative disks can be used to explain
the observations of the low/hard state in black hole binaries 
(for a review, see Remillard \& McClintock 2006). 
Here, we briefly discuss basic moments of this application of MAD's and leave
more quantitative considerations for future works.
We assume, for example, the development of the MAD in the
radiation pressure dominated accretion disk
at the subcritical regime (see Shakura \& Sunyaev 1973). 
In such a disk, the radiation diffusion time scale 
$t_{rad}\simeq H^2\sigma_T\rho/c$
can significantly exceed the Keplerian time
$t_{K}= 2\pi R^{3/2}/\sqrt{GM}$ at small values of the $\alpha$-parameter, $\alpha \la 0.1$,
in virtu of the relation
\begin{equation}
{t_{K}\over t_{rad}} \simeq 3.4\alpha,
\end{equation}
which follows from the Shakura-Sunyaev solution.
Here, we denote $H$ to be the disk half-thickness and $\sigma_T=0.4$ cm$^2$/g to be
the Thomson scattering cross-section.
As soon as $t_{rad}\gg t_{K}$ in the outer Shakura-Sunyaev disk,
$t_{rad}$ can also significantly exceed the accretion velocity $t_{accr}\sim t_{K}$
in the inner spiral flow in the MAD, in virtu of the relation
\begin{equation}
{t_{accr}\over t_{rad}}\propto R,
\end{equation}
which is satisfied in accretion flows with the scaling law of the accretion velocity
$v_{accr}\propto R^{-1/2}$ and $H\propto R$.
Having $t_{rad}\gg t_{accr}$, one concludes that the radiation is traped 
inside the spiral flow on the accretion time scale and, therefore, 
this flow is radiatively inefficient.
From the point of view of an observer, which detects
the softer part of the spectrum of outgoing radiation (below $\sim$ eVs),
the MAD will look like a Shakura-Sunyaev disk truncated at the inner radius
$R_{tr}$, which coinsides with the transition radius between the inner
magnetically dominated region and outer axisymmetric accretion disk. 
Typically, in observations, $R_{tr}$ is in the interval from a few tens 
to hundreds of $R_g$ (e.g., in Cyg X-1, see Done \& Zycki 1999), which
is consistent with that obtained in our Models~A and B.
The observed specta of black hole binaries in the low/hard state 
are dominated by the hard x-ray component
(e.g., Done, Gierli\'nski, \& Kubota 2007) and this
can be explained by the radiation from
the hot, optically thin magnetized medium, which surrounds the accreting spiral flows
and in which the binding energy of these flows is released.
The synchrotron radiation from the magnetized medium and Poynting jets in the MAD
(see Goldston, Quataert, \& Igumenshchev 2005) 
can be used to explain the observed radio luminosity
in the low/hard state; this luminosity is believed to be due to steady jets
(e.g., Corbel et al. 2003; Gallo, Fender, \& Pooley 2003).

Our simulations assume accretion disks
around black holes and can be relevant to objects with
relativistic jets containing accreting
stellar-mass black holes (e.g., in micro-quasars; black holes resulted from
type Ib/c supernova explosions and mergers of two compact objects) and
supermassive black holes (in galactic centers).
However, we believe that our main results
can also be relevant to nonrelativistic
astrophysical objects, in which accretion disks and jets are observed.
These objects include, for example, young stellar objects
and accreting stars (e.g., white dwarfs) in binary systems.
Qualitatively, we expect that the structure and dynamics of MAD's and
Poynting jets are similar in the both relativistic and nonrelativistic cases.
We expect, however, large quantitative differences
in these two cases
because of the different energy-density scales involved
in the regions of jet formation.
Black holes, which are capable of launching jets almost from
the event horizon, can produce ultra relativistic jets 
(e.g., McKinney 2006).
Jets from nonrelativistic objects, 
which have the surface radius $R_*\gg R_g$,
are limited by the velocities $v\sim\sqrt{GM/R_*}\ll c$.
For example, the latter formula gives the upper estimate of the jet velocity
$\sim 400\,km/s$ 
from the solar-type stars.

The problem of inward transport and amplification of the vertical field 
in turbulent accretion disks was intensively discussed 
in past and recent years (see, e.g., Spruit \& Uzdensky 2005).
The solution of this problem can help to discriminate models of
accretion disks, which are consistent with observations 
(e.g., Meier \& Nakamura 2006; Schild, Leiter, \& Robertson 2006).
Our simulation results show that the vertical field is transported inward
and amplified
independent of the disk structure, either laminar or turbulent.
It is worth noting, however, 
that magnetic fields in our models are imposed and
relatively large. The assumed strength of these fields 
exceeds the possible strength of the self-sustained magnetic fields that
could be developed due to the MRI (Sano et al. 2004).
Therefore, these results should be considered with some caution, because
they do not represent the case
of weak vertical magnetic fields.

The strong vertical field in the center of accretion disks
can be, in principal, 
a relic field that is inherited from the previous evolution. 
This field can appear, for example, in the merger scenario
(merger of two magnetized neutron stars or
a black hole with a magnetized neutron star, e.g., Berger et al. 2005) or
in the course of
the gravitational collapse of an extended ``proto" object 
(e.g., proto-stellar cloud, supernova progenitor),
which produces a significantly more compact object 
(protostar, black hole).
In the latter case,
the proto object may contain some amount of the poloidal field,
which will be amplified and accumulated at the center during the collapse.
After the formation of the compact object, the remained
noncollapsed mass can still
move inward, forming an accretion disk and confining
the field in the vicinity of the object.
Depending on the relative strength of this relic field
and the mass accretion rate, MAD's and Poynting jets can be developed.
The considered scenario can be applied
to young stellar objects (T-Tauri stars, e.g., Donati et al. 2007)
and the hyper-accretion model for gamma-ray bursts
(Woosley 1993; Paczy\'nski 1998).

The spiral-flow pattern in MAD's rotates with about
the same angular velocity at all radial distances; i.e.,
it rotates almost as a rigid body.
Such a rotation can result in
quasi-periodic oscillations (QPO's)
in the emitted radiation, if the disk's axis is inclined
to the line of view of an observer 
(e.g., Alpar \& Shaham 1985; Lamb et al. 1985; Strohmayer et al. 1996; 
Lamb \& Miller 2001; Titarchuk 2003).
The frequency of these QPO's should be related to the 
rotation of the spiral pattern,
which angular velocity is defined by the radius
of the magnetically dominated region and can be a
fraction ($\sim 0.5-1$) of the orbital velocity at this radius.
More investigations are required to make quantitative
predictions about QPO's from MAD's.

\acknowledgments

This work was supported by
the U.S. Department of Energy (DOE) Office of Inertial Confinement
Fusion under Cooperative Agreement No. DE-FC52-92SF19460, the
University of Rochester, the New York State Energy Research and
Development Authority.

\clearpage

\begin{figure}
\epsscale{.60}
\plotone{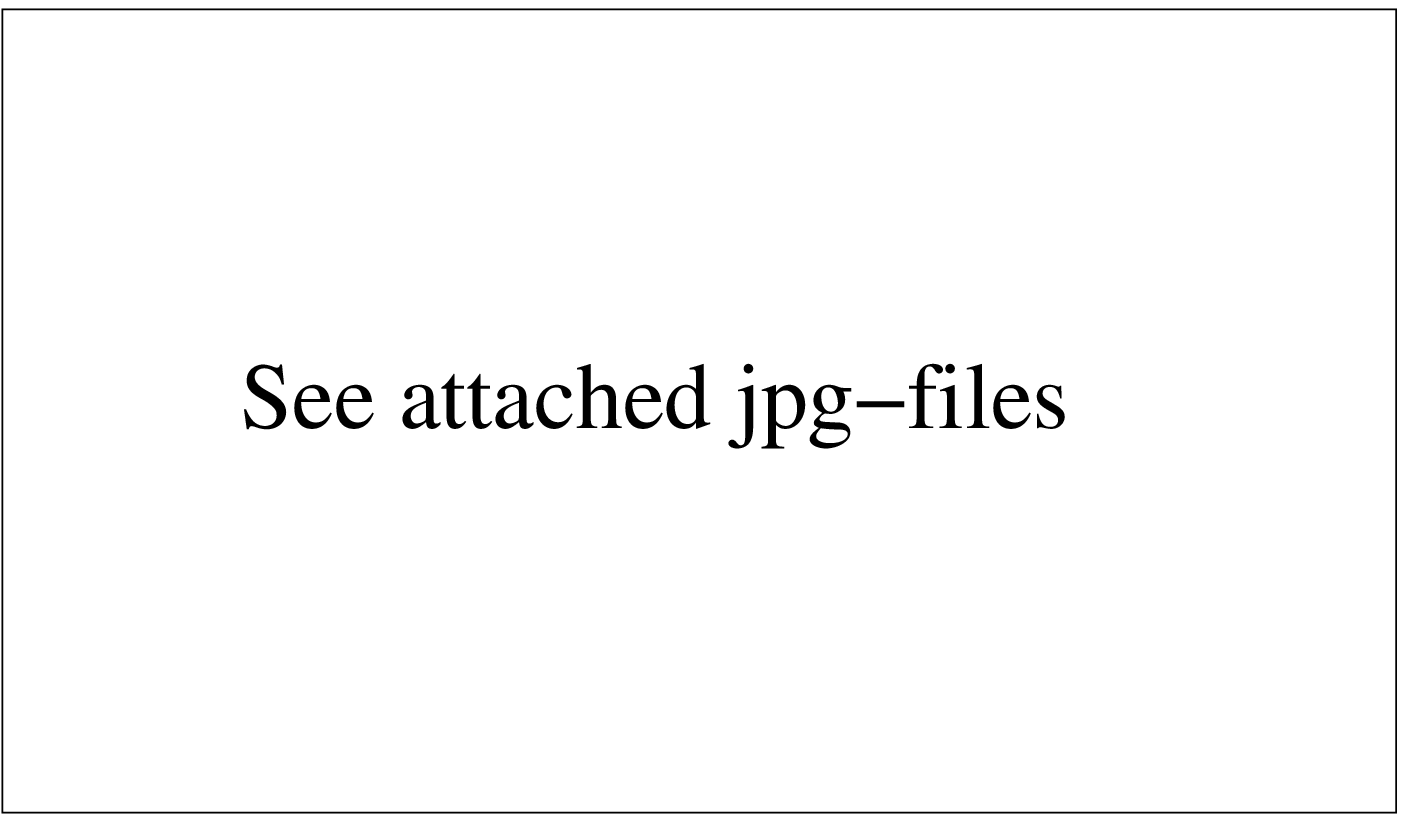}
\caption{Distributions of density in the meridional plane in Model~B
at two successive moments
(a) $t=5.1153$ and (b) $t=5.1458$ from axisymmetric 2D simulations. 
The black hole is located on the left and 
the small empty circle there corresponds to the inner
boundary around the black hole at $R_{in}=2\,R_g$. 
The axis of rotation is in the vertical direction.
The domain shown has the radial extent $100\,R_g$ along the equatorial plane
and represents a fraction of the full computational domain
with $R_{out}=220\, R_g$.
The time is in units of the orbital period at $R_{inj}=210\,R_g$.
The color bars on the bottom indicate the scales for $\log\rho$ 
(in arbitrary units).
The model develops the thick turbulent accretion disk, which is seen
as the nonuniform mass concentration near the equator.
The low-density polar regions are filled with the strong vertical
magnetic field. This field causes the cycle accretion into the black hole.
(a) The halted accretion period
when the mass is accumulated at the magnetospheric radius
$R_m\approx 15\,R_g$ and (b) the accretion period.
\label{fig1}}
\end{figure}

                                                                                
\begin{figure}
\epsscale{.60}
\caption{Snapshot of magnetic lines in Model~B
at the same moments and on the same spatial scales as in Fig.~1.
The component of the field lines parallel to the meridional plane
is shown. The accretion disk transports the vertical magnetic
flux inward, which accumulates in the vicinity of the black hole.
Small-scale magnetic loops are the result of turbulent motions
in the disk and disk corona.
(a) The halted accretion period (see Fig.~1a), in which
most of the accumulated magnetic flux
is outside the black hole horizon.
(b) The accretion period (see Fig.~1b), in which
all the accumulated flux goes through the horizon.
The lines have been plotted using the method of Cabral \& Leedom (1993).
\label{fig2}}
\end{figure}


\begin{figure}
\epsscale{.60}
\caption{Evolution of mass accretion rate and magnetic fluxes 
in Model~B. 
The time is in units of the orbital period at $R_{inj}$.
The simulations are performed in 2D, assuming axisymmetry,
from $t=0$ to 2.14 (the latter moment indicated by the vertical dashed line) 
and in 3D after that.
(a) The accretion into the black hole begins at $t\approx 1.3$.
Starting from $t\approx 1.4$, the cycle accretion is developed
(seen as a sequence of spikes). 
In 3D, the cycle accretion disappears.
(b) Magnetic fluxes (in arbitrary units) through the equatorial plane
inside five fixed radii: $2\,R_g$ ($=R_{in}$), $25\,R_g$, $50\,R_g$, 
$100\,R_g$, and $210\,R_g$ ($=R_{inj}$). The variability of the flux
at $R_{in}$ is related to the variability of the accretion rate in (a).
\label{fig3}}
\end{figure}

                                                                                
\begin{figure}
\epsscale{.60}
\caption{Distributions of density in the equatorial plane
in Model~B at two successive moments
(a) $t=2.2767$ and (b) $t=2.2867$ from 3D simulations.
The black hole is located in the center and the central 
black circle corresponds to the inner boundary at $R_{in}=2\,R_g$.
The shown domain has the extension 100 by $100\,R_g$.
The time is in units of the orbital period at $R_{inj}=210\,R_g$.
The color bars on the bottom indicate the scales for $\log\rho$
(in arbitrary units).
The accretion flow rotates in the clockwise direction and
forms the spiral structure inside $R\approx 35\,R_g$.
The low-density regions correspond to the magnetically dominated islands,
through which the vertical field penetrates the disk.
(a) The developed structure of spiral density inflows and
magnetic islands stretched in the azimuthal direction.
(b) The moment after about half a revolution of the spiral
magnetic pattern seen in (a); new magnetic islands inside $R\approx 15\,R_g$
are the result of an
eruption of the vertical magnetic flux from the inner boundary.
\label{fig4}}
\end{figure}

                                                                                
\begin{figure}
\epsscale{.60}
\caption{Evolution of mass accretion rate and magnetic fluxes
in Model~A. The notations used are the same as in Fig.~3.
Note the earlier beginning of accretion,
significantly larger accretion rate and magnetic fluxes, 
and longer quasi-periods between the accretion events
in the 2D simulations
in comparison with those in Model~B (see Fig.~3).
These are the result of a stronger field injection in Model~A.
\label{fig5}}
\end{figure}

                                                                                
\begin{figure}
\epsscale{.60}
\caption{Distribution of density in the equatorial plane
in Model~A at $t=1.82$ from 3D simulations.
The notations used are the same as in Fig.~4.
The shown domain has the extension 300 by $300\,R_g$.
The model is characterized by the relatively large magnetic flux accumulated
at the center. The magnetically dominated region has the outer
radius $R\simeq 70\,R_g$.
The accretion flow demonstrates the one-arm 
($m=1-$ mode) spiral structure in the innermost region. 
This structure rotates in the clock-wise direction and is split onto the three 
arms at larger radial distances.
There are other two large density streams in the process of development
(below the center and on the left from the center).
\label{fig6}}
\end{figure}


\begin{figure}
\epsscale{.60}
\caption{Distribution of Poynting $S_R$ (solid lines) and kinetic
$F_R$ (dashed lines) radial fluxes in the $\theta$-direction in Model~B.
The fluxes are averaged in the azimuthal direction and in time
over the interval $\Delta t=0.05$, beginning from $t=2.27$, and
normalized to the accretion power $\dot{M}_{in}c^2$.
The distributions are shown at the radial distances 
5 $R_g$, 10 $R_g$, 25 $R_g$, 50 $R_g$, 100 $R_g$, and 220 $R_g$ 
($=R_{out}$) in the panels from (a) to (f), respectively, and
illustrate the collimation of Poynting flux into bi-polar jets.
\label{fig7}}
\end{figure}


\begin{figure}
\epsscale{.60}
\caption{Same as in Fig.~7, but for Model~A.
The fluxes are averaged in time over the interval $\Delta t=0.04$, 
beginning from $t=1.78$. 
\label{fig8}}
\end{figure}


\begin{figure}
\epsscale{.60}
\caption{Radial distribution of Poynting jet
$\dot{E}_{jet}$ (solid line) and total $\dot{E}_{tot}$
(dashed line) luminosities in Model~B. The luminosities are normalized
to the accretion power $\dot{M}_{in}c^2$ and averaged in time
as explained in the capture to Fig.7.
\label{fig9}}
\end{figure}


\begin{figure}
\epsscale{.60}
\caption{Same as in Fig.~9, but for Model~A.
The luminosities are averaged in time
as explained in the capture to Fig.8.
\label{fig10}}
\end{figure}


\clearpage

\begin{thebibliography}{999}

\bibitem[]{} Agapitou, V., \& Papaloizou, J.C.B. 1996, 
  Astrophys. Lett. Commun., 34, 363
\bibitem[]{} Alpar, M. A., \& Shaham, J. 1985, Nature, 316, 239
\bibitem[]{} Appl, S. 1996, A\&A, 314, 995
\bibitem[]{} Balbus, S. A., \& Hawley, J. F. 1991, \apj, 376, 214
\bibitem[]{} Berger, E. et al. 2005, Nature, 438, 988
\bibitem[]{} Bisnovatyi-Kogan, G. S., \& Ruzmaikin, A. A. 1974, Ap\&SS, 28, 45
\bibitem[]{} Bisnovatyi-Kogan, G. S., \& Ruzmaikin, A. A. 1976, Ap\&SS, 42, 401
\bibitem[]{} Bisnovatyi-Kogan, G. S., \& Lovelace, R. V. E. 2007, \apj, 667, L167
\bibitem[]{} Blandford, R.D., \& Znajek, R.L. 1977, \mnras, 179, 433
\bibitem[]{} Blandford, R.D., \& Payne, D.G. 1982, \mnras, 199,883
\bibitem[]{} Brandenburg, A., Nordlund, A., Stein, R.F., \& Torkelsson, U.
  1995, \apj, 446, 741
\bibitem[]{} Cabral, B., \& Leedom, L. 1993, Computer Graphics: Proceedings:
  Annual Conference Series 1993: SIGGRAPH 93 (New York: Association for
  Computing Machinery), 263
\bibitem[]{} Chandran, B. D. G. 2001, \apj, 562, 737
\bibitem[]{} Colella, P., \& Woodward, P.R. 1984, J. Comp. Phys., 54, 174
\bibitem[]{} Corbel, S., Nowak, M. A., Fender, R. P., Tzioumis, A. K., \& Markoff, S.
  2003, Astron. Astrophys., 400, 1007
\bibitem[]{} De Villiers J.-P., Hawley, J. F., \& Krolik, J. H. 
  2003, \apj, 599, 1238
\bibitem[]{} Donati, J.-F., Jardine, M. M., Gregory S. G., Petit, P.,
  Bouvier, J., Dougados, C., M\'enard, F., Cameron, A. C., Harries, T. J.,
  Jeffers, S. V., \& Paletou, F. 2007, \mnras, 380, 1297
\bibitem[]{} Done, C., \& Zycki, P. T. 1999, \mnras, 305, 457
\bibitem[]{} Done, C., Gierli\'nski, M., \& Kubota, A. 2007, Astron. Astrophys. Rev., in press
\bibitem[]{} Eichler, D. 1993, \apj, 419, 111
\bibitem[]{} Evans, C.R., \& Hawley, J.F. 1988, \apj, 332, 659
\bibitem[]{} Gallo, E., Fender, R. P., \& Pooley, G. G. 2003, \mnras, 344, 60
\bibitem[]{} Gardiner, T.A., \& Stone, J.M. 2005, J. Comp. Phys., 205, 509
\bibitem[]{} Goldston, J. E., Quataert, E., \& Igumenshchev, I. V. 2005, \apj, 621, 785
\bibitem[]{} Hawley, J. F., \& Balbus, S. A. 1992, \apj, 400, 595
\bibitem[]{} Hawley, J.F., Gammie, C.F., \& Balbus, S.A. 1996, \apj, 464, 690
\bibitem[]{} Hawley, J. F., \& Krolik, J. H. 2006, \apj, 641, 103
\bibitem[]{} Heyvaerts, J., Priest, E.R., \& Bardou, A. 1996, \apj, 473, 403
\bibitem[]{} Hirose, S., Krolik, J.H., De Villiers, J.-P., \& Hawley, J.F.
  2004, \apj, 606, 1083
\bibitem[]{} Igumenshchev, I. V. 2006, \apj, 649, 361
\bibitem[]{} Igumenshchev, I. V., Narayan, R., \& Abramowicz, M. A. 2003, 
  \apj, 592, 1042
\bibitem[]{} Kaisig, M., Tajima, T., \& Lovelace, R. V. E. 1992, \apj, 386, 83
\bibitem[]{} Kato, S., Fukue, J., \& Mineshige, S. 1998, Black-Hole
     Accretion Disks (Kyoto: Kyoto Univ. Press)
\bibitem[]{} Kato, Y., Mineshige, S., \& Shibata, K. 2004, \apj, 605, 307
\bibitem[]{} Komissarov, S.S. 2005, \mnras, 359, 801
\bibitem[]{} K\"onigl, A. 1989, \apj, 342, 208
\bibitem[]{} K\"onigl, A., \& Pudritz, R. E. 2000, 
  in Protostars and Planets IV, ed. V. Mannings, A. Boss, \& S. Russell 
  (Arizona: Univ. Arizona Press), p.759
(astro-ph/9903168)
\bibitem[]{} Lamb, F. K., \& Miller, M. C. 2001, \apj, 554, 1210
\bibitem[]{} Lamb, F. K., Shibazaki, N., Alpar, M. A., \& Shaham, J. 
  1985, Nature, 317, 681
\bibitem[]{} Landau, L. D., \& Lifshitz, E. M. 1987, Electrodynamics of
  Continuous Media. Pergamon Press, Oxford
\bibitem[]{} Li, L.-X. 2000, \apj, 531, L111
\bibitem[]{} Li, L.-X., \& Narayan, R. 2004, \apj, 601, 414
\bibitem[]{} Li, S. 2005, J. Comp. Phys., 203, 344
\bibitem[]{} Livio, M., Ogilvie, G.I., \& Pringle, J.E. 1999, \apj, 512, 100
\bibitem[]{} Livio, M., Pringle, J.E., \& King, A.R. 2003, \apj, 593, 184
\bibitem[]{} Lovelace, R. V. E., Wang, J. C. L., \& Sulkanen, M. E. 1987,
   \apj, 315, 504
\bibitem[]{} Lovelace, R. V. E., Li, H., Koldoba, A. V., Ustyugova, G. V., \&
   Romanova, M. M. 2002, \apj, 572, 445
\bibitem[]{} Lovelace, R. V. E., Gandhi, P. R., \& Romanova, M. M. 2005,
  Ap\&SS, 298, 115
\bibitem[]{} Lovelace, R. V. E., Romanova, M. M., \& Newman, W. I. 1994,
  \apj, 437, 136
\bibitem[]{} Lubow, S.H., Papaloizou, J.C.B., \& Pringle, J.E. 1994,
   \mnras, 267, 235
\bibitem[]{} Lynden-Bell, D. 2003, \mnras, 341, 1360
\bibitem[]{} Meier, D.L., Nakamura, M. 2006, ASPC, 350, 195
\bibitem[]{} McKinney, J. C., \& Gammie, C. F. 2004, \apj, 611, 977
\bibitem[]{} McKinney, J. C. 2006, \mnras, 368, 1561
\bibitem[]{} Narayan, R., Igumenshchev, I. V., \& Abramowicz, M. A.
     2000, \apj, 539, 798
\bibitem[]{} Narayan, R., Igumenshchev, I. V., \& Abramowicz, M. A.
     2003, PASJ, 55, L69
\bibitem[]{} Paczy\'nski, B., \& Wiita, P.J. 1980, A\&A, 88, 23
\bibitem[]{} Paczy\'nski, B. 1998, \apj, 494, L45
\bibitem[]{} Proga, D., \& Zhang, B. 2006, \mnras, 370, L61
\bibitem[]{} Punsly, B. 2001, Black Hole Gravitohydromagnetics 
  (New York: Springer)
\bibitem[]{} Quataert, E., \& Gruzinov, A. 2000, \apj, 539, 809
\bibitem[]{} Remillard, R. A., \& McClintock, J. E. 2006, Annu. Rev. Astron Astrophys., 44, 49
\bibitem[]{} Romanova, M. M., \& Lovelace, R. V. E. 2006, \apj, 645, L73
\bibitem[]{} Sano, T., Inutsuka, S., Turner, N.J., Stone, J.M. 2004,
  \apj, 605, 321
\bibitem[]{} Schild, R.E., Leiter, D.J., \& Robertson, S.L. 2006, \aj, 132, 420
\bibitem[]{} Shakura, N. I., \& Sunyaev, R.A. 1973, A\&A, 24, 337
\bibitem[]{} Spruit, H.C., Foglizzo, T., \& Stehle, R. 1997, \mnras, 288, 333
\bibitem[]{} Spruit, H. C., Stehle, R., \& Papaloizou, J. C. B. 1995,
  \mnras, 275, 1223
\bibitem[]{} Spruit, H.C., \& Uzdensky, D.A. 2005, \apj, 629, 960
\bibitem[]{} Stone, J.M., Hawley, J.F., Gammie, C.F., \& Balbus, S.A. 1996,
  \apj, 463, 656
\bibitem[]{} Stone, J. M., \& Pringle, J. E. 2001, \mnras, 322, 461
\bibitem[]{} Stone, J. M., \& Norman, M. L. 1994, \apj, 433, 746
\bibitem[]{} Strohmayer, T. E., Zhang, W., Swank, J. H., Smale, A., 
  Titarchuk, L., Day, C., \& Lee, U. 1996, \apj, 469, L9
\bibitem[]{} Takahashi, M., Nitta, S., Tatematsu, Y., \&
  Tomimatsu, A. 1990, \apj, 363, 206
\bibitem[]{} Titarchuk, L. 2003, \apj, 591, 354
\bibitem[]{} Tomimatsu, A., Matsuoka, T., \& Takahashi, M. 2001,
  Phys. Rev. D, 64, 123003
\bibitem[]{} Tout, C. A., \& Pringle, J. E. 1996, \mnras, 281, 219
\bibitem[]{} van Ballegooijen, A.A. 1989, in Proc. European Phys. Soc. 
  Study Conf., Accretion Disks and Magnetic Fields in Astrophysics,
  ed. G. Belvedere (Dordrecht: Kluwer), 99
\bibitem[]{} Woosley, S.E. 1993, \apj, 405, 273

\end{thebibliography}
\end{document}